\def\beg{\begin{equation}}
\def\eeq{\end{equation}}
\documentstyle[12pt]{article}
\begin{document}
\begin{center}
{\Large{\bf Comments on ``Scattering of bunched fractionally charged quasiparticles" by Chung, Heiblum and Umansky, cond-mat/0305325.}}
\vskip0.35cm
{\bf Keshav N. Shrivastava}
\vskip0.25cm
{\it School of Physics, University of Hyderabad,\\
Hyderabad  500046, India}
\end{center}

In the experiments, the quantity measured is the product of the charge and the magnetic field from which the fractional charge is deduced. There is no objection to measuring the fractional charge as long as it is remembered that the product of the charge and the field has been measured. So if the fraction came from the field rather than from charge, the experiment will remain unaffected. There is no 
prescription about mass splitting so there is no way to combine two fractionally charged quasiparticles into one. Therefore, the fractional charge can be obtained by changing the state of the quasiparticle without splitting, then there is no bunching.
\vskip1.0cm
Corresponding author: keshav@mailaps.org\\
Fax: +91-40-2301 0145.Phone: 2301 0811.
\vskip1.0cm

We have recently noted a paper by Chung et al[1] in which fractional charges are claimed to have been seen. This will create misleading impression on the theorists so that it is necessary to clarify the quantity being measured. It is reasonable to start with the paper of Laughlin[2] where it was first thought that a wave function for a fractionally charged excitation has been found.
 The usage of the terminology ``fractional charge" is perfectly justified but we should understand the quantity which is part of the problem. Laughlin found a wave function $|\psi_m|^2$, which describes a system uniformly expanded to a charge density of,
\beg
\rho={e\over m(2\pi a_o^2)}.
\eeq
It minimizes the energy when $\rho$ equals the charge density generating the potential. When we substitute $a_o$=1 and m= odd number such as 3,
then the charge density becomes $\rho = {e\over 3} ({1\over 2\pi})$
and we think that we are finding a charge of $e/3$. Since $a_o$=1 is a constant, at this time we can ignore it but we can keep in mind that the lengths must be constant or in other words, we have an ``incompressible" system. Then with this restriction, we do have a fractional charge in an incompressible system. Now, when we relax the incompressibility, the factor of 3 can be absorbed in $a_o$ and hence the charge of e/3 again becomes $e$. The  {\bf correct quantity } is $e/a_o^2$. Since flux is quantized,
\beg
Ba_o^2=n\phi_o
\eeq
or we can change the quantity $e/a_o^2$ to $eB/n\phi_o$. Hence, the quantity is not the charge but it is $eB$. The unit flux is $\phi_o$=hc/e so the charge density is,
\beg
\rho ={eB\over m 2\pi n\phi_o}.
\eeq
Therefore, instead of changing the charge, we can change some other quantity such as $B$. This problem has been discussed[3] in another eprint. It is sufficient to say that the experimentalist measures the product of the charge and field but not the charge. The experimental work of Goldman, Su, de-Picciotto, Reznikov, Banin, Saminadayar, Glattli, Jin, Etienne, Comforti, Chung, Heiblum, Umansky, Mahalu, etc has been examined in another eprint[4] in which the same conclusion is reached. Therefore, the experimentalists who claim to have measured the fractional charge have actually measured the product of the charge and the field but not charge alone.

     Chung et al also use the composite fermion model, according to which `` even number of flux quanta" are attached to the electron. In this connection, it is found[5] that many of the experimentally observed fractional charges are not in agreement with the CF model. A very extensive study shows that ``even number of flux quanta" are not attached to the electron and hence CF model should be discarded[6].

     The measurements were conducted by Chung et al[1] by setting the magnetic field within the conductance plateau. This resulted into measurement of field multiplied by charge and not charge alone. The expression used for the shot noise, worked out in absence of field,
but there is a field present is,
\beg
S=4k_BTg+2qI_Bt.\theta(T,V)
\eeq
measures the product $qI_B$. There is a large field present in the system. Therefore, the above formula should be reworked out with field present. Since $I_B$ is equivalent to a field, the product of the current and field is measured and not the charge alone. Here, the high voltage transmission is $t=g/g_Q$, the ratio of conductivity to the quantized conductivity, 
\beg
I_B=Vg_Q(1-t),
\eeq
and,
\beg
\theta(T,V)=coth(qV/2k_BT)-2k_BT/qV.
\eeq
In view of the approximations
\beg
S\simeq 2qI_B
\eeq
so that still charge is not alone. If a factor of 1/3 came, then we can not know whether it came from $q$ or from $I_B$. Chung et al convert the thermal noise power and conductance into temperature by using the relation $S=4k_BTg$ where $g$ is the conductance. They obtain, $T\simeq 9 mK$, for the back scattering potential strength and claim the voltage
and temperature dependence of the differential conductance to be positive in agreement with Luttinger liquid. In the case of Luttinger liquid, the boson and fermions become indistinguishable. However, if we take the well known expressions for the boson and the fermion distributions, they never cross. Therefore, the claim of Chung et al to find agreement in a real laboratory experiment is not justified. Chung et al say that the  shot noise is due to a charge $q=e/3$ at $T=9 mK$. It depends on V, q, t and T. It can as well be that the factor of 1/3 is not associated with the charge. Chung et al use the CF model to say that second Landau level is involved with $p=2$ and $e/3$. Three quasiparticles of charge $e/7$ each make a quasiparticle of charge $3e/7$ and two quasiparticles of charge $e/5$ each make one quasiparticle of charge $2e/5$ at $\simeq 9 mK$. If this is correct,
what makes the components attractive? Two quasiparticles of charge $e/5$ each are likely to be ${\it repulsive}$ so they will not bunch to make one of charge $2e/5$.

     If three quasiparticles of charge $e/3$ each bunch then a quasiparticle of charge $e$ is made. Is this the correct method of making the electrons? Laughlin does not attach flux quanta to electrons but CF attach flux quanta to electrons. Luttinger mixes the bosons and fermion statistices. Chung et al are therefore using internally incompatible ideas and measurements of fractional charges, even if correct, are not based on ``fractionalization of charge".

     If the mass of a quasiparticle is ${\sl m}/3$ and charge $e/3$
then charge over mass ratio is $e/{\sl m}$ and the factor of 1/3 disappears. If electron splits into three particles of mass ${\sl m}/3$
each, then such a mass should appear in the formulas but there is no such expression. Then there are beautiful graphs showing e/3. The correct interpretation is that the quasiparticle of charge e/3 is seen in a magnetic field and then in a different event a quasiparticle
of charge $e$ is seen but $e$ is not made by bunching three quasiparticles of charge $e/3$ each with no prescription for mass.
Similarly, the charge $2e/5$ need not be made by bunching of two
quasiparticles of charge $1e/5$ each. The number of particles in
the quasiparticle of charge $2/5$ is only one. A single particle can change its charge by changing its state so there is no need of
combining two particles.

   It may be that the state changes by changing ${\it l}$ and
$s$ and then the Lande's splitting factor becomes 2 so that the 
magnetic moment becomes $2\mu_B$ which
is equivalent to changing $e$ to $2e$, but the number of electrons is only one. Thus $2e$ can arise in one electron. Similarly, there may be other combinations of ${\it l}$ and $s$ which give various values
of the fractional charges for only ${\it one}$ electron.

     Chung et al have not found any prescription for mass splitting and hence the idea of bunching of two fractionally charged quasiparticles to form one quasiparticle is not supported by the theory. If quasiparticles of fractional charge have to bunch together to make a quasiparticle of double the charge, then there must be a process which splits one quasiparticle into two quasiparticles. It may be that there is always only one particle, the effective charge of which changes as it goes from one level to anther. So different charges can be observed without splitting or bunching. We have found[7,8] that in the case of quantum Hall effect various fractional charges can be made by changing the state, such as ${\it l}$ and $s$, of a particle.
     
\noindent{\bf~~ Conclusions}.

     In conclusion, we find that charge measurements are actually measurements of the product of charge and the magnetic field. The various amounts of effective charge can be acquired by one particle without bunching.

\vskip1.25cm

\noindent{\bf ~References}
\begin{enumerate}
\item Y. C. Chung, M. Heiblum and V.Umansky, cond-mat/0305325.
\item R. B. Laughlin, Phys. Rev. Lett. {\bf 50}, 1395(1983).
\item K. N. Shrivastava, Cond-mat/0210238.
\item K. N. Shrivastava, Cond-mat/0302461.
\item Pan et al, Phys. Rev. Lett. {\bf 90}, 016801 (2003).
\item K. N. Shrivastava, Cond-mat/0304269;
 cond-mat/0304014 and 0302610
\item K. N. Shrivastava, Cond-mat/0303309.
\item K. N. Shrivastava, Introduction to quantum Hall effect,\\
      Nova Science Pub. Inc., N. Y. (2002).
\end{enumerate}
\vskip0.1cm

\end{document}